\documentclass{elsart}
\usepackage{graphicx}
\begin{document} 
\begin{frontmatter}
\title{Investigation of Galilean Invariance of multi-phase lattice Boltzmann methods}
\author{ A.J. Wagner \and Q. Li} 
\address{Department of Physics, North Dakota State University, Fargo 58105, ND}

\begin{abstract}
We examine the Galilean invariance of standard lattice Boltzmann
methods for two-phase fluids. We show that the known
Galilean invariant term that is cubic in the velocities, and is
usually neglected, is the main source of Galilean invariance
violations. We show that incorporating a correction term can improve
the Galilean invariance of the method by up to an order of
magnitude. Surprisingly incorporating this correction term can also
noticeably increase the range of stability for multi-phase
algorithms. We found that this is true for methods in which the
non-ideality is incorporated by a forcing term as well as methods in
which non-ideality is directly incorporated in a non-ideal pressure
tensor.
\end{abstract}


\end{frontmatter}

\section{INTRODUCTION}
The issue of Galilean invariance for lattice Boltzmann methods, and in
particular for non-ideal fluid simulations based on an input pressure
tensor\cite{swift}, has received considerable attention in the
past\cite{Holdych,Inamuro,Kalarakis}. It was noticed that these
methods had a severe problems with Galilean invariance and careful
expansion methods elucidated a set of correction terms for the
pressure tensor to improve Galilean invariance. The attraction of
methods based on an input pressure tensor is that this pressure
tensor can be easily derived from an input free energy. This
immediately delivers predictions for the phase-diagram, the interface
profiles and the surface tension.

Others held that the problem lay deeper, and that it was inappropriate
to include the non-ideal terms for the method in an input pressure
tensor, and that it is appropriate to put the non-ideal terms in a
Vlasov-like forcing term\cite{Shan,Luo}. Unfortunately, it has been
difficult to relate the approach by Shan and Chen\cite{Shan} to a free
energy. Meanwhile others have taken to use a forcing term that
corresponds to the divergence of the pressure tensor\cite{Luo}. This should, in
principle, lead to an equivalent model to the pressure tensor
approach. We will show later, that there are subtle differences
though. On the up-side, using a forcing approach does not
require similar correction terms to the pressure tensor approach, and
therefore it has been labeled ``Galilean invariant''.

Recently we have been working on including Lees-Edwards-like boundary
conditions in lattice Boltzmann \cite{LE} to simulate sheared
systems. It turns out that these boundary conditions are very
sensitive to Galilean invariance violations. None of the above
methods seemed to show a sufficient level of Galilean invariance which
lead us to this closer investigation of the origins of Galilean
invariance in lattice Boltzmann methods.

In a lattice based method the lattice represents a fixed reference
frame, and it is not surprising that this should show up to some
order. So we measured the error for a variety of situations and tried
to quantify the main contribution. We found that the well known
$\nabla^2(\rho u^3)$ term in the Taylor expansion of the momentum
equations, which is usually neglected, is responsible for the majority
of the error. We then demonstrate how including an additional
correction term can significantly increase the Galilean invariance of
lattice Boltzmann methods.

\section{The lattice Boltzmann method} \label{secLB}
We can write the lattice Boltzmann method in a general way that neatly
separates the ideal gas contributions from the non-ideal
contributions. To do this we write
\begin{equation} 
f_i({\bf r} + {\bf v} \Delta t, t+\Delta t)-f_i({\bf r},t)=\frac{1}{\tau}(f_i^0({\bf x},t)-f_i({\bf x},t)+G_i)+F_i \label{LBE}
\label{LB}
\end{equation}
where $f^0_i$ is the contribution function for an ideal gas, the
$F_i$ are the contributions of a forcing term and the $G_i$ allow us
to manipulate the pressure tensor. As usual the moments for the
ideal-gas distribution function are
\begin{eqnarray*}
\sum_i f_i^0=\rho,\;\;
\sum f_i^0 v_{i\alpha}=\rho \tilde{u}_\alpha,\;\; 
\sum_if_i^0 v_{i\alpha}v_{i\beta}=c_s^2 \rho\delta_{\alpha\beta} 
+\rho \tilde{u}_\alpha \tilde{u}_\beta,\\
\sum_i f^0_i v_{i\alpha} v_{i\beta} v_{i\gamma}=
\frac{1}{3} \rho (\tilde{u}_\gamma\delta_{\alpha\beta} + \tilde{u}_\beta
\delta_{\alpha\gamma} + \tilde{u}_\alpha \delta_{\beta \gamma})+\rho \tilde{u}_\alpha
\tilde{u}_\beta \tilde{u}_\gamma + Q_{\alpha \beta \gamma}
\end{eqnarray*}
where $\rho=\sum_i f_i$ is the mass density, $\tilde{u}=\sum_i f_i v_i / \rho $ is the
mean fluid velocity before the action of the forcing term $F_i$. The
velocity of sound is given by $c_s=1/\sqrt{3}$. For all models with
$v_{i\alpha}^2=1$ (like D1Q3, D2Q7, D2Q9, D3Q15, D3Q19 or D3Q27) we have
$v_{i\alpha}^3 = v_{i\alpha}$ which means that \textit{e.g.} $\sum_i f_i
v_{ix}^3=\sum_i f_i v_{ix}=\rho u_x$ making a correction term $Q$ necessary
for the third moment. One usually chooses $ Q_{\alpha \beta \gamma} = - \rho u_\alpha u_\beta u_\gamma
$. It is this term that will lead to the leading Galilean invariance
problems.

To simulate fluids with a non-ideal equation of state we can introduce
either a forcing term $F_i$ with the moments
\begin{eqnarray*}
&&\sum_i F_i= 0,\;\;
\sum F_i v_{i\alpha}=\rho a_\alpha,\;\;  
\sum F_i v_{i\alpha} v_{i\beta} = \rho (a_\alpha u_\beta + a_\beta
u_\alpha),\\
&& \sum F_i v_{i\alpha} v_{i\beta}v_{i\gamma} =
\frac{1}{3}(\rho a_\alpha \delta_{\beta \gamma} + \rho a_\beta \delta_{\alpha
\beta} + \rho a_\gamma \delta_{\alpha \beta})
\end{eqnarray*} 
or a pressure term $G_i$ with the moments
\begin{eqnarray*}
\sum G_i = 0,\;\; 
\sum G_i v_{i\alpha}=0,\;\;
\sum_i G_i v_{i\alpha}v_{i\beta}=A_{\alpha \beta},\;\;
\sum_i G v_{i\alpha} v_{i\beta}v_{i\gamma} =0.  
\end{eqnarray*}
Here $\rho a$ is a forcing term and $A$ is a pressure term. We will see
below how these terms can be used to introduce non-ideal terms into
the equation of state or simply to correct the above mentioned
deficiencies of the velocity set.

The knowledge of these moments is sufficient to perform a Taylor
expansion (or equivalently a Chapman Enskog multi-scale expansion) of
equation (\ref{LB}). If we define the macroscopic velocity $u$ as
$u=\tilde{u}+a/2$ we obtain the continuity equation
\begin{equation}
\partial_t \rho +\partial_\alpha (\rho u_\alpha)=0
\label{cont}
\end{equation}
and a momentum conservation equation
\begin{eqnarray}
\partial_t (\rho u_\alpha) + \partial_\beta(\rho u_\alpha
u_\beta)
&=& - \partial_\beta (\rho c_s^2\delta_{\alpha\beta}+ A_{\alpha
  \beta}) 
+ \rho a_\alpha \nonumber\\
&&+  \partial_\beta [\nu\rho (\partial_\beta u_\alpha + \partial_\alpha u_\beta
+ \partial_\gamma u_\gamma \delta_{\alpha \beta})] \label{NS}\\ & & -
\nu \partial_\beta [ u_\alpha \partial_\gamma A_{\beta \gamma} +
u_\beta \partial_\gamma A_{\alpha \gamma}+ \partial_\rho A_{\alpha \beta} \partial_\gamma (\rho
u_\gamma)+\partial_\gamma Q_{\alpha
\beta \gamma} ]\nonumber
\end{eqnarray}
where the kinematic viscosity is $\nu=(\tau-1/2)c_s^2$. This equation is
the Navier-Stokes equation, except for the terms in the last line. The
condition of Galilean invariance requires that a description in a
reference frame $S$ another reference frame $S'$ translating with a
constant velocity $u_0$ be related by a spatial coordinate
transformation $x=x'-u_0 t$ and a translation of velocities
$u=u'-u_0$. It is easy to see that only the terms in the last line of
eqn. (\ref{NS}) are not Galilean invariant.

Let us first consider (\ref{NS}) for an ideal gas. In this case
$a=0$ and $A=0$ so the
only non-Galilean invariant term to second order is $Q$. We can
eliminate this error term by introducing a well crafted forcing term
of the form
\begin{equation}
\rho a_\alpha = \nu \partial_\beta \partial_\gamma Q_{\alpha\beta\gamma}
\label{corr}
\end{equation}
We should mention here that we can avoid this problem and choose
$Q=0$ if we use a velocity set that is large enough\cite{icase}. But
for now we want to stay with the standard velocity sets.

To simulate a non-ideal system we want to obtain a pressure term in
the first line of (\ref{NS}) that is the divergence of a pressure
tensor derived from Thermodynamics. In particular we need
\begin{equation}
\partial_\beta P_{\alpha\beta} = \partial_\beta(c_s^2 \rho
\delta_{\alpha\beta} +A_{\alpha\beta})+\rho a_\alpha
\end{equation}
to first order. For the forcing approach we choose $A=0$ and this
equation defines $a$. We will refer to this approach as ``Forcing'' in
our comparisons. To improve Galilean invariance for this approach we
can add the same additional forcing term (\ref{corr}) to restore
Galilean invariance to second order. This approach we will refer to as
``ForcingQ''.

If we chose to introduce the non-ideal term in $A$ we can recover the
algorithm of Swift \textit{et al.}\cite{swift}. This corresponds to
choosing $A_{\alpha\beta}=P_{\alpha\beta}-\rho c_s^2 $ and $a=0$. We refer to this
approach in the following as ``Pressure''. 

This approach is deficient in as much as the correction terms in
the third line in (\ref{NS}) are unphysical and will lead to severely
non-Galilean invariant behavior. This problem was addressed
by Holdych \textit{et al.}\cite{Holdych} and
independently by Inamuro \textit{et al.}\cite{Inamuro}. Because the
divergence of the pressure tensor is small to first order, the main
contribution to the error terms comes from the density
alone. Therefore choosing
\begin{equation}
A_{\alpha\beta}=P_{\alpha\beta}-\rho c_s^2 \rho \delta_{\alpha\beta}
-\nu (\partial_\alpha \rho u_\beta+\partial_\beta \rho
u_\alpha+\partial_\gamma \rho u_\gamma)
\end{equation}
will only leave error terms of the order $\partial^2 P$ which can be
assumed to be small in systems close to equilibrium. This approach
will be referred to as ``Holdych''. The restriction of being close
to equilibrium was late revisited by Kalarakis \textit{et
  al.}\cite{Kalarakis} who suggested an improvement to these corrections.

However, the Holdych approach as well as all other later approaches
are still deficient in that they do not correct the Q term. Clearly
there is a multitude of possible combinations of choices for $A$ and
$a$ that will lead to a Galilean invariant form of
eqn. (\ref{NS}). One other choice we examined is
\begin{eqnarray*}
A_{\alpha\beta}&=&P_{\alpha\beta}-\rho c_s^2,\\
a_\alpha&=&- \partial_\beta\{\nu \partial_\gamma [u_\alpha (P_{\beta\gamma}-\rho c_s^2\delta_{\beta\gamma})+u_\beta (A_{\alpha\gamma}-\rho c_s^2\delta_{\alpha\gamma})
+(\partial_\rho A_{\alpha\beta}-c_s^2\delta{\alpha\beta})\partial_\gamma(\rho u_\gamma)]\}.
\end{eqnarray*}
which has in common with the Kalarakis\cite{Kalarakis} approach that
it is not limited to systems close to equilibrium.  The above choice
leads to a Galilean invariant momentum equation given by
\begin{eqnarray}
\partial_t (\rho u_\alpha) + \partial_\beta(\rho u_\alpha
u_\beta)
&=& - \partial_\beta (\rho c_s^2\delta_{\alpha\beta}+ A_{\alpha
  \beta}) 
+ \rho a_\alpha \nonumber\\
&&+  \partial_\beta [\nu (P_{\beta\gamma} \partial_\gamma u_\alpha + P_{\alpha\gamma} \partial_\gamma u_\beta
+ P_{\alpha\beta}\partial_\gamma u_\gamma))] \label{NS2}
\end{eqnarray}
This scheme has a tensorial interface viscosity.  We refer to the
resulting method as ``PressureQ''.

\section{The non-ideal gas} 
For simplicity we will consider a non-ideal gas with a $\phi^4$-free
energy\cite{Briant}. For such a system we can calculate the phase
diagram and the surface tension analytically, simplifying the analysis. 

For a critical density $\rho_c$, critical temperature $T_c$ and
critical pressure $p_c$ we obtain for the pressure tensor
\begin{equation}
P_{\alpha \beta} =[p_c (\phi + 1) ^2 (3 \phi^2 - 2 \phi + 1 + 2
\theta ) - \kappa \rho \nabla^2 \rho - \frac{\kappa}{2} (\nabla \rho)^2
]\delta_{\alpha \beta} + \kappa\partial_\alpha \rho\partial_\beta
\rho
\end{equation} 
 where $\phi = (\rho-\rho_c)/\rho_c$ is the reduced density and $\theta =\beta (T - T_c)
/T_c $ is the reduced temperature, where $\beta$ is an arbitrary constant.
The equilibrium values for the density are given by
\begin{equation}
 \rho^0 = \rho_c\pm \sqrt{- \theta }.
\end{equation}
The definition of the pressure tensor is all that is needed to
define the lattice Boltzmann methods for non-ideal fluids as we
explained in section \ref{secLB}.

\begin{figure}
\begin{center}
\resizebox{0.5\textwidth}{!}{\includegraphics{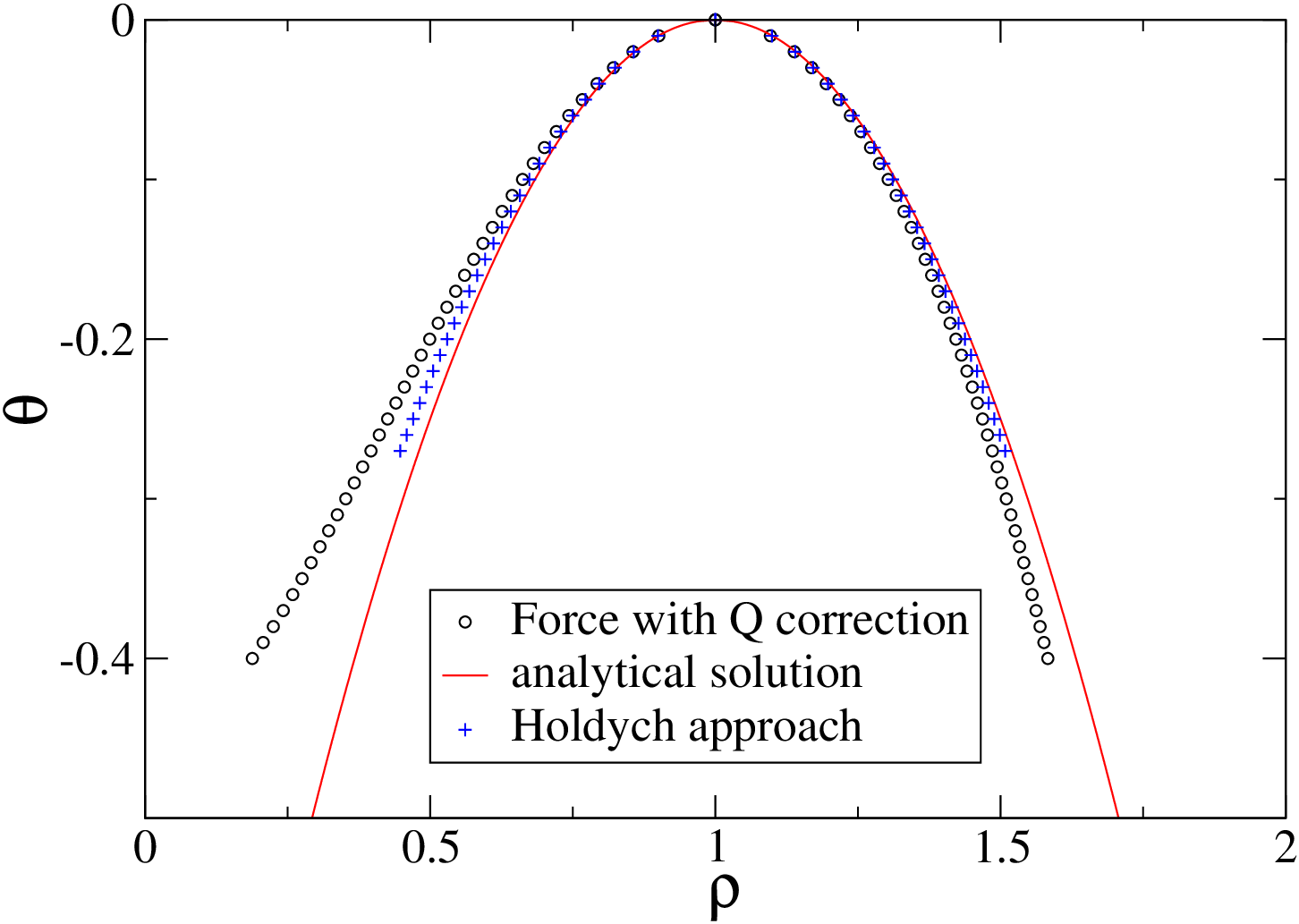}} 
\end{center}
\caption{ Phase diagram for $p_c$=0.1, $\kappa$ = 0.1, $n_c$ = 1,$p_c$= 0.1
  and $\nu=1/6$ for the ``ForcingQ'' and ``Holdych'' approaches. Note
  that the Holdych approach is better at reproducing the analytical
  phase diagram where as the ForcingQ approach has a larger range of
  stability.}
\label{phasediagram}
\end{figure}

We performed simulations of an equilibrium system containing one
domain of gas and one of liquid at different imposed velocities.
As mentioned above, while the different approaches
are very similar as far as the expansion to second order is concerned,
there are noticeable differences in the behavior of the methods.
 Firstly let us compare the numerical results for the
phase-diagram shown in Figure \ref{phasediagram}. On the one hand we
notice that the ability of the pressure based methods to reproduce the
analytical phase-diagram is noticeably better than the corresponding
forcing method. On the other hand we we see that the range of
stability for the forcing method is larger, leading to a larger
maximum density ratio (of about six) which can be simulated with this
method. 

\begin{figure}
\begin{center}
\resizebox{0.5\textwidth}{!}{\rotatebox{0}{\includegraphics{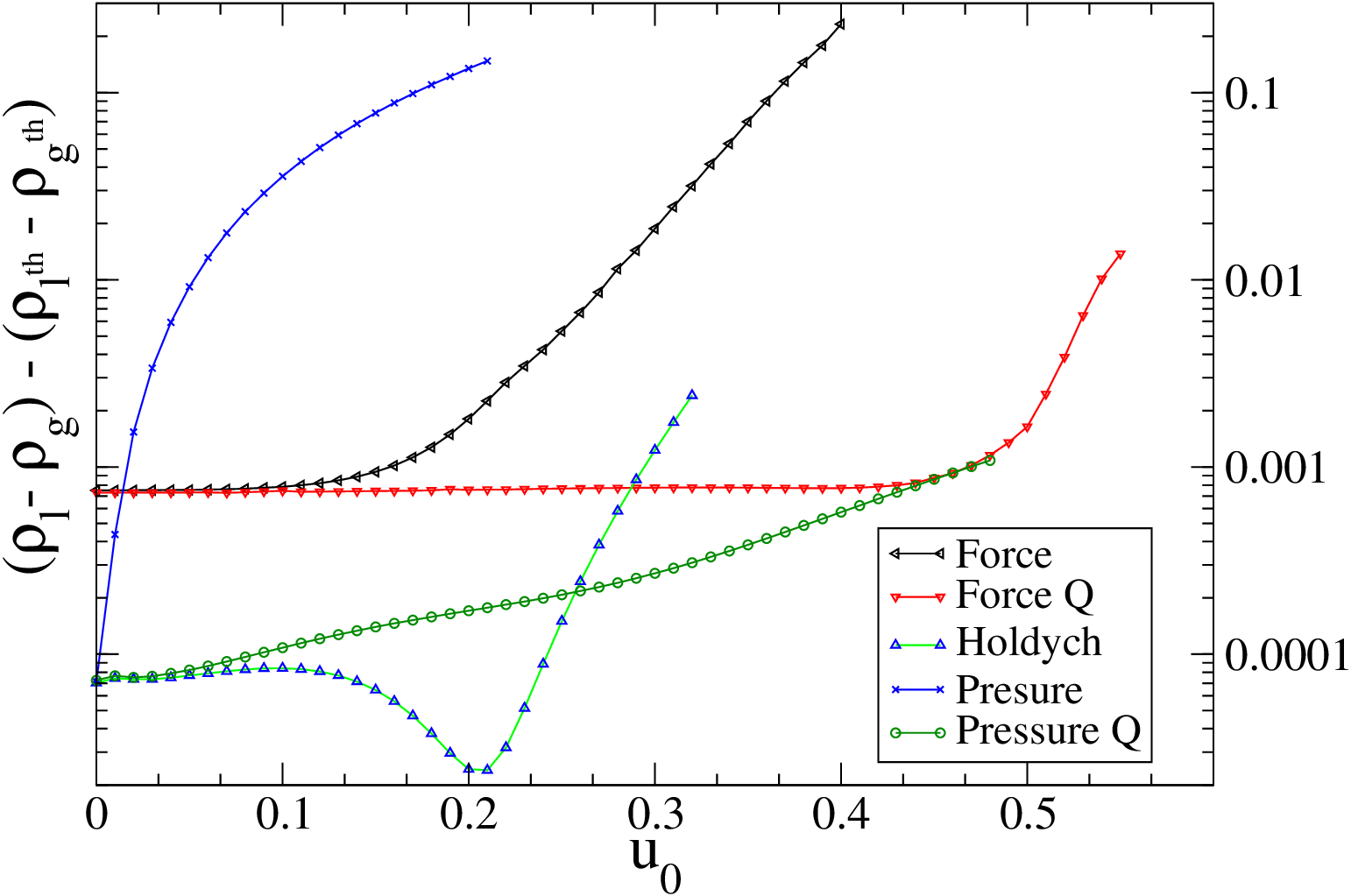}}} 
\end{center}
\caption{Deviation from the analytical density difference for the
  different methods as a function of $u0$. $\kappa$ = 0.1, $n_c$ = 1.0, $p_c$ = 0.42, 
 $\beta = 0.1$, $\theta=-0.03$ and $\nu =1/6$. The data end at values of $u_0$ at
  which the methods became unstable.} 
\label{Figdev}
\end{figure}

Next we examine the effect of an imposed velocity on the equilibrium
densities. In Figure \ref{Figdev} we show the deviation of the
predicted liquid-gas density difference from the measured one. We
notice that the original pressure method has excessive deviations even
for small $u_o$. We were surprised to see that in these simulations the gas
and liquid domains were actually stationary, even with an imposed
velocity $u_0$. The domains made up for the imposed velocity by
evaporation and condensation mechanisms and a faster velocity in the
gas than in the fluid. We see that the error is generally less for the
corrected pressure methods than for the forcing methods.

We were very surprised to see that the $Q$ corrections did not only
improve that Galilean invariance of the method, they substantially
increased the range of stability. This is true both for the range of
stable velocities $u_0$ as well as the accessible density ratios (even
at $u_0=0$. This was an unexpected benefit of our study and we still
do not understand why the correction term has such beneficial effects
on the stability.

We now need to quantify the Galilean invariance error for the
advection of an interface profile. For the velocity
the analytical solution is a constant, so we can define an error function
\begin{equation}
E_D(u_0)=\sqrt{\sum_x (u(x)-u_0)^2/L_x}.
\end{equation}
This measure is effectively time independent, as is appropriate for
this equilibrium consideration.

\begin{figure}
\begin{center}
\resizebox{0.5\textwidth}{!}{\rotatebox{0}{\includegraphics{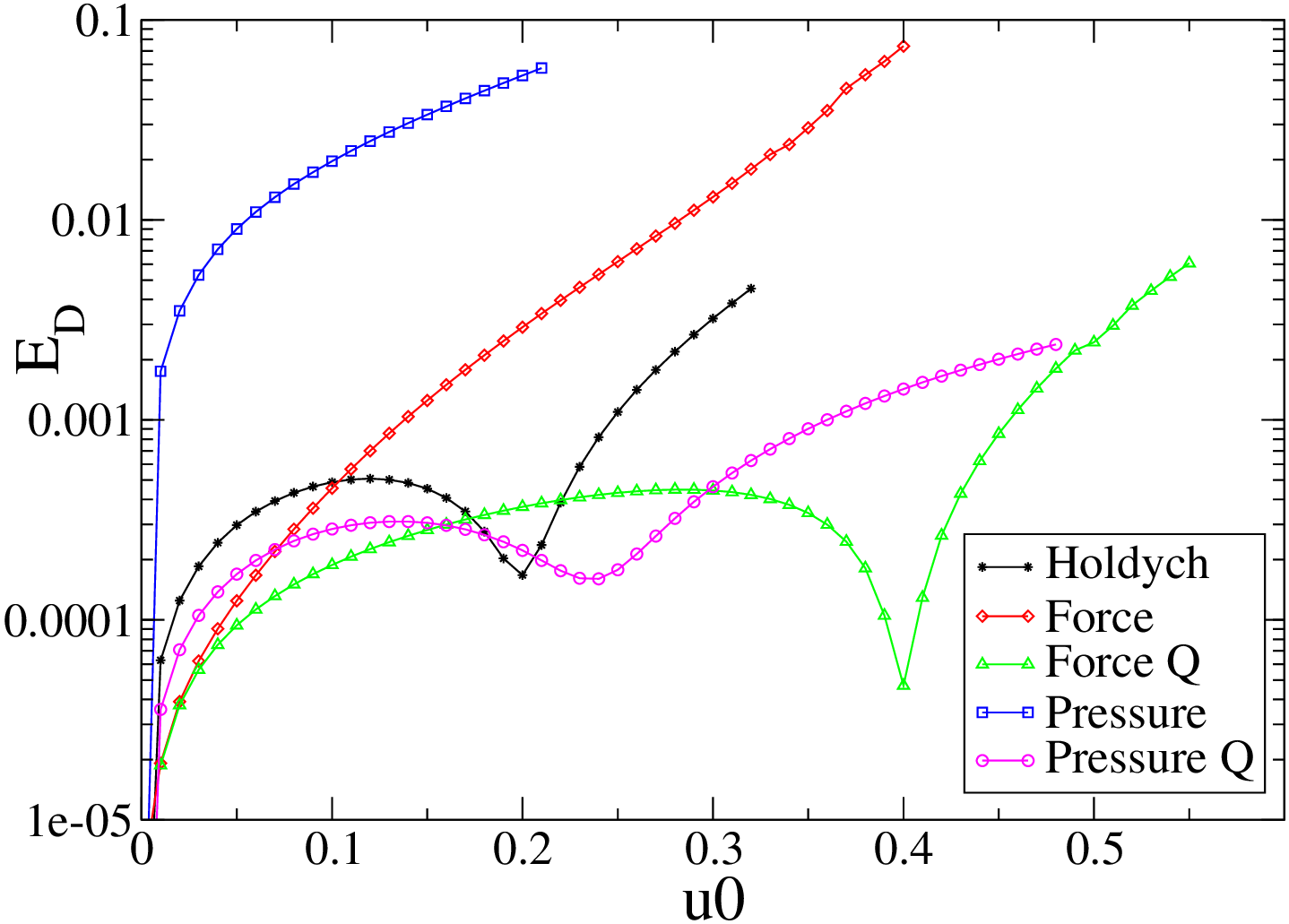}}}
\end{center}
\caption{$E_D(u_0)$ for simulation parameters $\kappa=0.01$, $n_c=1$, $p_c=0.42$,
 $\theta=-0.03$, iteration=10000 for the different methods.}
\label{figPhaseGal}
\end{figure} 

In Figure \ref{figPhaseGal} we show the Galilean invariance error
$E_D(u_o)$ for the different methods we described. As mentioned above
the original pressure approach performs very poorly, in fact refusing
to advect the domains relative to the lattice. The Holdych approach
improves on this significantly leading to an advection of the
profile. We were surprised by the non-monotonic behavior of the error,
leading to a minimum at $u_0=0.2$. We will come back to this
later. Adding the $Q$ correction to the Holdych approach leads to a
noticeable improvement for velocities as small as $u_0=10^{-3}$. The
Forcing approach leads to a good behavior at small $u_0$ but increases
rapidly with $u_0$. Its behavior is significantly improved for
$u_0>0.03$ by including the $Q$ correction in the in the method.

\begin{figure}
\begin{center}
\resizebox{0.5\textwidth}{!}{\rotatebox{0}{\includegraphics{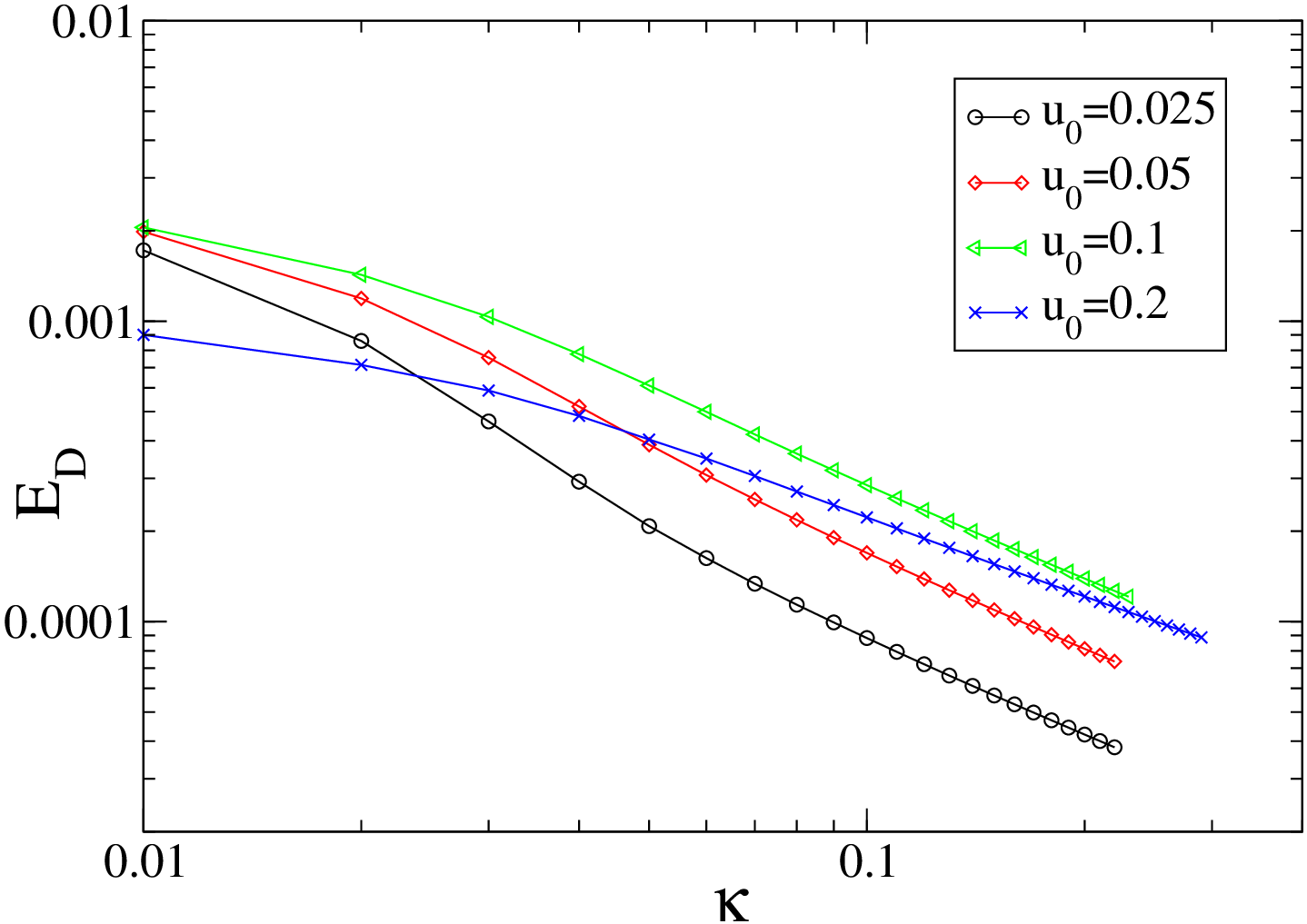}}}
\end{center}
\caption{$E_D(u_0)$ as a function of $\kappa$ for different values for
  $u_0$ for the PressureQ method. Other parameters are $n_c=1$,
  $p_c=0.42$, $\theta=-0.03$.}
\label{figKappadep}
\end{figure} 

When interpreting the above results, it is important to remember that
the parameter space for the Galilean invariance problem includes not
only $u_0$ but also the parameters determining the equilibrium density
profile $\kappa$, $\theta$ and $p_c$ as well as the relaxation time $\tau$. This
parameter space is so large as to make it nearly impossible to examine
it exhaustively. But we want to discuss at least the dependence on
$\kappa$, which is related to the interface width and the surface
tension. It is also important to look at this when one wants to fairly
compare the pressure an forcing approaches. It turns out that while
the pressure approach reproduces the analytical interface profile
fairly, the forcing approach does not, at least not for the nominal
value of $\kappa$. The forcing approach leads to a much wider interface so
that it would be fairer to compare the forcing approach to a pressure
approach with a larger $\kappa$. In Figure \ref{figKappadep} we can see
that this matters a lot. The Galilean invariance error $E_D$ is
largest for small values of $\kappa$ corresponding to thin interfaces. And
for small values of $\kappa$ this absolute error decreases much slower with
decreasing $u_0$. This is
probably the reason that the pressure approach performed much worse
than the forcing approach (with the same nominal $\kappa$ value) for small
$u_0$.

A close examination of Fig \ref{figKappadep} also shows that the
non-monotonic behavior of $E_D(u_0)$ is related to the small $\kappa$
behavior. The graph suggests that the error function becomes monotonic
for large $\kappa$.

\section{Summary}
We have shown that the usually neglected error term in the Navier
Stokes level momentum equation derived for standard lattice Boltzmann
methods leads to noticeable Galilean invariance violations. These
violations are noticeable even in the range of usual velocities of
$u<0.1$, but become dominant for larger velocities.

A carefully defined forcing term can remove the non-Galilean invariant
terms recovering the Navier Stokes equation. We have shown that this
approach is also effective in practice, reducing the Galilean invariance
error substantially. The correction term also had the added benefit of
increasing the range of stability for the multi-phase applications,
leading to a larger range of stable velocities and even, perhaps
surprisingly, to a larger range of density ratios that can be
simulated by the methods.

\end{document}